%Paper: hep-th/9311026
%From: bern@physics.ucla.edu (Bern)
%Date: Wed, 3 Nov 93 19:54:32 PST

%%%%%%%%%%%%%%%%%%%%%%%%%%%%%%%%%%%%%%%%%%%%%%%%%%%%%%%%%%%%%%%%%%%%%%%%%%%%%
%                  New QCD Results from String Theory                       %
%                    Z. Bern, L. Dixon, D.A. Kosower                        %
%      Talk presented by Z.B. at Strings 1993, May 24-29, Berkeley CA       %
%                       To Appear in Proceedings                            %
%%%%%%%%%%%%%%%%%%%%%%%%%%%%%%%%%%%%%%%%%%%%%%%%%%%%%%%%%%%%%%%%%%%%%%%%%%%%%

\magnification=\magstep1

\newbox\SlashedBox
\def\slashed#1{\setbox\SlashedBox=\hbox{#1}
\hbox to 0pt{\hbox to 1\wd\SlashedBox{\hfil/\hfil}\hss}#1}
\def\hboxtosizeof#1#2{\setbox\SlashedBox=\hbox{#1}
\hbox to 1\wd\SlashedBox{#2}}

% The following is necessary so that we can get a partial slash
% inside a math display... sigh.
\def\mathslashed#1{\setbox\SlashedBox=\hbox{$#1$}
\hbox to 0pt{\hbox to 1\wd\SlashedBox{\hfil/\hfil}\hss}#1}

\def\ifsmall{\iffalse}  % default is unreduced.
\def\titlepagefont{}  % default is ordinary font.

% the ps: landscape must be the first special command in order
% to get the first page in landscape mode -- so we go through some
% contortions to define TeXgraphics in the default case.
\def\DefineTeXgraphics{%
\special{ps::[global] /TeXgraphics { } def}}  % No need to do anything

\def\today{\ifcase\month\or January\or February\or March\or April\or May
\or June\or July\or August\or September\or October\or November\or
December\fi\space\number\day, \number\year}
\def\eatPrefix19{}
\def\Year{\expandafter\eatPrefix\the\year}
\newcount\hours \newcount\minutes
\def\monthname{\ifcase\month\or
January\or February\or March\or April\or May\or June\or July\or
August\or September\or October\or November\or December\fi}
\def\shortmonthname{\ifcase\month\or
Jan\or Feb\or Mar\or Apr\or May\or Jun\or Jul\or
Aug\or Sep\or Oct\or Nov\or Dec\fi}

\def\TimeStamp{\hours\the\time\divide\hours by60%
\minutes -\the\time\divide\minutes by60\multiply\minutes by60%
\advance\minutes by\the\time%
${\rm \shortmonthname}\cdot\if\day<10{}0\fi\the\day\cdot\the\year%
\qquad\the\hours:\if\minutes<10{}0\fi\the\minutes$}

%\DefineTeXgraphics}

%\DefineTeXgraphics}

%\DefineTeXgraphics}

%\DefineTeXgraphics}

% restores pagenumbers

%\def\draft{\centerline{\it Preliminary Draft}\vskip 0.4in}

\newif\ifdraftmode
\newif\ifleftlabels  % Labels in left margins as well, for European-size paper

% Stolen from harvmac.tex 04/08/92
%       use \nolabels to get rid of eqn, ref, and fig labels in draft mode
\def\nolabels{\def\wrlabeL##1{}\def\eqlabeL##1{}\def\reflabeL##1{}}
\def\writelabels{\def\wrlabeL##1{\leavevmode\vadjust{\rlap{\smash%
{\line{{\escapechar=` \hfill\rlap{\sevenrm\hskip.03in\string##1}}}}}}}%
\def\eqlabeL##1{{\escapechar-1\rlap{\sevenrm\hskip.05in\string##1}}}%
\def\reflabeL##1{\noexpand\rlap{\noexpand\sevenrm[\string##1]}}}
\def\writeleftlabels{\def\wrlabeL##1{\leavevmode\vadjust{\rlap{\smash%
{\line{{\escapechar=` \hfill\rlap{\sevenrm\hskip.03in\string##1}}}}}}}%
\def\eqlabeL##1{{\escapechar-1%
\rlap{\sixrm\hskip.05in\string##1}%
\llap{\sevenrm\string##1\hskip.03in\hbox to \hsize{}}}}%
\def\reflabeL##1{\noexpand\rlap{\noexpand\sevenrm[\string##1]}}}
\nolabels

\newdimen\fullhsize
\newdimen\hstitle
\hstitle=\hsize % default
\newdimen\hsbody
\hsbody=\hsize % default
\newdimen\hbodyoffset
\hbodyoffset=\hoffset % default
\newbox\leftpage
\def\abstract#1{#1}
\def\rotated{\special{ps: landscape}
\magnification=1000  % This line must come before we change vsize,
                     % since \magnification sets it to a fixed value.
\baselineskip=14pt
\global\hstitle=9truein\global\hsbody=4.75truein
\global\vsize=7truein\global\voffset=-.31truein
\global\hoffset=-0.54in\global\hbodyoffset=-.54truein
\global\fullhsize=10truein
\def\DefineTeXgraphics{%
\special{ps::[global]
/TeXgraphics {currentpoint translate 0.7 0.7 scale
              -80 0.72 mul -1000 0.72 mul translate} def}}
 % 0.7 is slightly less than the ratio of horizontal sizes: 4.75 to 6.5
\let\lr=L
\def\ifsmall{\iftrue}
\def\titlepagefont{\twelvepoint}
\trueseventeenpoint
\def\almostshipout##1{\if L\lr \count1=1
      \global\setbox\leftpage=##1 \global\let\lr=R
   \else \count1=2
      \shipout\vbox{\hbox to\fullhsize{\box\leftpage\hfil##1}}
      \global\let\lr=L\fi}

\output={\ifnum\count0=1 %%% This is the HUTP version
 \shipout\vbox{\hbox to \fullhsize{\hfill\pagebody\hfill}}\advancepageno
 \else
 \almostshipout{\leftline{\vbox{\pagebody\makefootline}}}\advancepageno
 \fi}

\def\abstract##1{{\leftskip=1.5in\rightskip=1.5in ##1\par}} }

% Messages on lines by themselves
\def\linemessage#1{\immediate\write16{#1}}

% tagged sec numbers
\global\newcount\secno \global\secno=0
\global\newcount\appno \global\appno=0
\global\newcount\meqno \global\meqno=1
\global\newcount\subsecno \global\subsecno=0
% and figure numbers
\global\newcount\figno \global\figno=0

\newif\ifAnyCounterChanged
% If we are comparing numbers, there's no special problem.
% But if we are comparing roman numerals, we must be careful, because
% stuff read in from the aux file would be made up of ordinary
% characters (category code = 11), whereas \romannumeral generates
% characters with category code = 12..., so the stuff from the
% current run won't appear equal to the previous definition, as far
% as \warnIfChanged is concerned.
% To get around this, we have a macro \makeNormal, which converts
% letters `ivxlcdmIVXLCDM' to normal letters, no matter what their category
% code.  The macro has the convoluted form it does, with aftergroup's & all,
% to avoid blowing up TeX...
% The macro is used below in makeNormalizedRomappno, by which means we
% define the appendix counters to be strings containing vanilla versions
% of the letters... Sigh
\let\terminator=\relax
% The string to be normalized must not contain { and } tokens...
\def\normalize#1{\ifx#1\terminator\let\next=\relax\else%
\if#1i\aftergroup i\else\if#1v\aftergroup v\else\if#1x\aftergroup x%
\else\if#1l\aftergroup l\else\if#1c\aftergroup c\else%
\if#1m\aftergroup m\else%
\if#1I\aftergroup I\else\if#1V\aftergroup V\else\if#1X\aftergroup X%
\else\if#1L\aftergroup L\else\if#1C\aftergroup C\else%
\if#1M\aftergroup M\else\aftergroup#1\fi\fi\fi\fi\fi\fi\fi\fi\fi\fi\fi\fi%
\let\next=\normalize\fi%
\next}
% makes #1 a normalized version of #2...
\def\makeNormal#1#2{\def\doNormalDef{\edef#1}\begingroup%
\aftergroup\doNormalDef\aftergroup{\normalize#2\terminator\aftergroup}%
\endgroup}
% makes a normalized version of its argument:

\def\warnIfChanged#1#2{%
\ifundef#1% skip it
\else\begingroup%
\edef\oldDefinitionOfCounter{#1}\edef\newDefinitionOfCounter{#2}%
%\message{old: \oldDefinitionOfCounter}%
%\message{new: \newDefinitionOfCounter}%
\ifx\oldDefinitionOfCounter\newDefinitionOfCounter%
\else%
\linemessage{Warning: definition of \noexpand#1 has changed.}%
\global\AnyCounterChangedtrue\fi\endgroup\fi}

\def\Section#1{\global\advance\secno by1\relax\global\meqno=1%
\global\subsecno=0%
\bigbreak\bigskip% (combination \goodbreak\bigskip\bigskip)
\centerline{\twelvepoint \bf %
\the\secno. #1}%
\par\nobreak\medskip\nobreak}
\def\tagsection#1{%
\warnIfChanged#1{\the\secno}%
\xdef#1{\the\secno}%
\ifWritingAuxFile\immediate\write\auxfile{\noexpand\xdef\noexpand#1{#1}}\fi%
}
\def\section{\Section}
\def\Subsection#1{\global\advance\subsecno by1\relax\medskip %
\leftline{\bf\the\secno.\the\subsecno\ #1}%
\par\nobreak\smallskip\nobreak}
\def\tagsubsection#1{%
\warnIfChanged#1{\the\secno.\the\subsecno}%
\xdef#1{\the\secno.\the\subsecno}%
\ifWritingAuxFile\immediate\write\auxfile{\noexpand\xdef\noexpand#1{#1}}\fi%
}

\def\subsection{\Subsection}

\def\romappno{\uppercase\expandafter{\romannumeral\appno}}
\def\makeNormalizedRomappno{%
\expandafter\makeNormal\expandafter\normalizedromappno%
\expandafter{\romannumeral\appno}%
\edef\normalizedromappno{\uppercase{\normalizedromappno}}}
\def\Appendix#1{\global\advance\appno by1\relax\global\meqno=1\global\secno=0
\bigbreak\bigskip% (combination \goodbreak\bigskip\bigskip)
\centerline{\twelvepoint \bf Appendix %
\romappno. #1}%
\par\nobreak\medskip\nobreak}
\def\tagappendix#1{\makeNormalizedRomappno%
\warnIfChanged#1{\normalizedromappno}%
\xdef#1{\normalizedromappno}%
\ifWritingAuxFile\immediate\write\auxfile{\noexpand\xdef\noexpand#1{#1}}\fi%
}
\def\appendix{\Appendix}

\def\eqn#1{\makeNormalizedRomappno%
\ifnum\secno>0%
  \warnIfChanged#1{\the\secno.\the\meqno}%
  \eqno(\the\secno.\the\meqno)\xdef#1{\the\secno.\the\meqno}%
     \global\advance\meqno by1
\else\ifnum\appno>0%
  \warnIfChanged#1{\normalizedromappno.\the\meqno}%
  \eqno({\rm\romappno}.\the\meqno)%
      \xdef#1{\normalizedromappno.\the\meqno}%
     \global\advance\meqno by1
\else%
  \warnIfChanged#1{\the\meqno}%
  \eqno(\the\meqno)\xdef#1{\the\meqno}%
     \global\advance\meqno by1
\fi\fi%
\eqlabeL#1%
\ifWritingAuxFile\immediate\write\auxfile{\noexpand\xdef\noexpand#1{#1}}\fi%
}
\def\defeqn#1{\makeNormalizedRomappno%
\ifnum\secno>0%
  \warnIfChanged#1{\the\secno.\the\meqno}%
  \xdef#1{\the\secno.\the\meqno}%
     \global\advance\meqno by1
\else\ifnum\appno>0%
  \warnIfChanged#1{\normalizedromappno.\the\meqno}%
  \xdef#1{\normalizedromappno.\the\meqno}%
     \global\advance\meqno by1
\else%
  \warnIfChanged#1{\the\meqno}%
  \xdef#1{\the\meqno}%
     \global\advance\meqno by1
\fi\fi%
\eqlabeL#1%
\ifWritingAuxFile\immediate\write\auxfile{\noexpand\xdef\noexpand#1{#1}}\fi%
}
\def\anoneqn{\makeNormalizedRomappno%
\ifnum\secno>0
  \eqno(\the\secno.\the\meqno)%
     \global\advance\meqno by1
\else\ifnum\appno>0
  \eqno({\rm\normalizedromappno}.\the\meqno)%
     \global\advance\meqno by1
\else
  \eqno(\the\meqno)%
     \global\advance\meqno by1
\fi\fi%
}
\def\mfig#1#2{\global\advance\figno by1%
\relax#1\the\figno%
\warnIfChanged#2{\the\figno}%
\edef#2{\the\figno}%
\reflabeL#2%
\ifWritingAuxFile\immediate\write\auxfile{\noexpand\xdef\noexpand#2{#2}}\fi%
}

\catcode`@=11 % borrow the private macros of PLAIN (with care)

%\font\titlefont=cmr10 at 16pt
\font\ninerm=cmr9
\font\eightrm=cmr8
\font\sixrm=cmr6

\def\loadtrueseventeenpoint{
 \font\seventeenrm=cmr10 at 17.28truept
 \font\seventeeni=cmmi10 at 17.28truept
 \font\seventeenbf=cmbx10 at 17.28truept
 \font\seventeenit=cmti10 at 17.28truept
 \font\seventeensl=cmsl10 at 17.28truept
 \font\seventeensy=cmsy10 at 17.28truept
}
\def\loadfourteenpoint{
\font\fourteenrm=cmr10 at 14.4pt
\font\fourteeni=cmmi10 at 14.4pt
\font\fourteenit=cmti10 at 14.4pt
\font\fourteensl=cmsl10 at 14.4pt
\font\fourteensy=cmsy10 at 14.4pt
\font\fourteenbf=cmbx10 at 14.4pt
}
\def\loadtruetwelvepoint{
\font\twelverm=cmr10 at 12truept
\font\twelvei=cmmi10 at 12truept
\font\twelveit=cmti10 at 12truept
\font\twelvesl=cmsl10 at 12truept
\font\twelvesy=cmsy10 at 12truept
\font\twelvebf=cmbx10 at 12truept
}

\font\ninei=cmmi9
\font\eighti=cmmi8
\font\sixi=cmmi6
\skewchar\ninei='177 \skewchar\eighti='177 \skewchar\sixi='177

\font\ninesy=cmsy9
\font\eightsy=cmsy8
\font\sixsy=cmsy6
\skewchar\ninesy='60 \skewchar\eightsy='60 \skewchar\sixsy='60

\font\ninebf=cmbx9
\font\eightbf=cmbx8
\font\sixbf=cmbx6

\font\ninett=cmtt9
\font\eighttt=cmtt8

\hyphenchar\tentt=-1 % inhibit hyphenation in typewriter type
\hyphenchar\ninett=-1
\hyphenchar\eighttt=-1

\font\ninesl=cmsl9
\font\eightsl=cmsl8

\font\nineit=cmti9
\font\eightit=cmti8

 % unslanted text italic

\newskip\ttglue
\def\tenpoint{\def\rm{\fam0\tenrm}%
  \textfont0=\tenrm \scriptfont0=\sevenrm \scriptscriptfont0=\fiverm
  \textfont1=\teni \scriptfont1=\seveni \scriptscriptfont1=\fivei
  \textfont2=\tensy \scriptfont2=\sevensy \scriptscriptfont2=\fivesy
  \textfont3=\tenex \scriptfont3=\tenex \scriptscriptfont3=\tenex
  \def\it{\fam\itfam\tenit}\textfont\itfam=\tenit
  \def\sl{\fam\slfam\tensl}\textfont\slfam=\tensl
  \def\bf{\fam\bffam\tenbf}\textfont\bffam=\tenbf \scriptfont\bffam=\sevenbf
  \scriptscriptfont\bffam=\fivebf
  \normalbaselineskip=12pt
  \let\sc=\eightrm
  \let\big=\tenbig
  \setbox\strutbox=\hbox{\vrule height8.5pt depth3.5pt width\z@}%
  \normalbaselines\rm}

\def\twelvepoint{\def\rm{\fam0\twelverm}%
  \textfont0=\twelverm \scriptfont0=\ninerm \scriptscriptfont0=\sevenrm
  \textfont1=\twelvei \scriptfont1=\ninei \scriptscriptfont1=\seveni
  \textfont2=\twelvesy \scriptfont2=\ninesy \scriptscriptfont2=\sevensy
  \textfont3=\tenex \scriptfont3=\tenex \scriptscriptfont3=\tenex
  \def\it{\fam\itfam\twelveit}\textfont\itfam=\twelveit
  \def\sl{\fam\slfam\twelvesl}\textfont\slfam=\twelvesl
  \def\bf{\fam\bffam\twelvebf}\textfont\bffam=\twelvebf
  \scriptfont\bffam=\ninebf
  \scriptscriptfont\bffam=\sevenbf
  \normalbaselineskip=12pt
  \let\sc=\eightrm
  \let\big=\tenbig
  \setbox\strutbox=\hbox{\vrule height8.5pt depth3.5pt width\z@}%
  \normalbaselines\rm}

\def\fourteenpoint{\def\rm{\fam0\fourteenrm}%
  \textfont0=\fourteenrm \scriptfont0=\tenrm \scriptscriptfont0=\sevenrm
  \textfont1=\fourteeni \scriptfont1=\teni \scriptscriptfont1=\seveni
  \textfont2=\fourteensy \scriptfont2=\tensy \scriptscriptfont2=\sevensy
  \textfont3=\tenex \scriptfont3=\tenex \scriptscriptfont3=\tenex
  \def\it{\fam\itfam\fourteenit}\textfont\itfam=\fourteenit
  \def\sl{\fam\slfam\fourteensl}\textfont\slfam=\fourteensl
  \def\bf{\fam\bffam\fourteenbf}\textfont\bffam=\fourteenbf%
  \scriptfont\bffam=\tenbf
  \scriptscriptfont\bffam=\sevenbf
  \normalbaselineskip=17pt
  \let\sc=\elevenrm
  \let\big=\tenbig
  \setbox\strutbox=\hbox{\vrule height8.5pt depth3.5pt width\z@}%
  \normalbaselines\rm}

\def\seventeenpoint{\def\rm{\fam0\seventeenrm}%
  \textfont0=\seventeenrm \scriptfont0=\fourteenrm \scriptscriptfont0=\tenrm
  \textfont1=\seventeeni \scriptfont1=\fourteeni \scriptscriptfont1=\teni
  \textfont2=\seventeensy \scriptfont2=\fourteensy \scriptscriptfont2=\tensy
  \textfont3=\tenex \scriptfont3=\tenex \scriptscriptfont3=\tenex
  \def\it{\fam\itfam\seventeenit}\textfont\itfam=\seventeenit
  \def\sl{\fam\slfam\seventeensl}\textfont\slfam=\seventeensl
  \def\bf{\fam\bffam\seventeenbf}\textfont\bffam=\seventeenbf%
  \scriptfont\bffam=\fourteenbf
  \scriptscriptfont\bffam=\twelvebf
  \normalbaselineskip=21pt
  \let\sc=\fourteenrm
  \let\big=\tenbig
  \setbox\strutbox=\hbox{\vrule height 12pt depth 6pt width\z@}%
  \normalbaselines\rm}

\def\ninepoint{\def\rm{\fam0\ninerm}%
  \textfont0=\ninerm \scriptfont0=\sixrm \scriptscriptfont0=\fiverm
  \textfont1=\ninei \scriptfont1=\sixi \scriptscriptfont1=\fivei
  \textfont2=\ninesy \scriptfont2=\sixsy \scriptscriptfont2=\fivesy
  \textfont3=\tenex \scriptfont3=\tenex \scriptscriptfont3=\tenex
  \def\it{\fam\itfam\nineit}\textfont\itfam=\nineit
  \def\sl{\fam\slfam\ninesl}\textfont\slfam=\ninesl
  \def\bf{\fam\bffam\ninebf}\textfont\bffam=\ninebf \scriptfont\bffam=\sixbf
  \scriptscriptfont\bffam=\fivebf
  \normalbaselineskip=11pt
  \let\sc=\sevenrm
  \let\big=\ninebig
  \setbox\strutbox=\hbox{\vrule height8pt depth3pt width\z@}%
  \normalbaselines\rm}

\def\eightpoint{\def\rm{\fam0\eightrm}%
  \textfont0=\eightrm \scriptfont0=\sixrm \scriptscriptfont0=\fiverm%
  \textfont1=\eighti \scriptfont1=\sixi \scriptscriptfont1=\fivei%
  \textfont2=\eightsy \scriptfont2=\sixsy \scriptscriptfont2=\fivesy%
  \textfont3=\tenex \scriptfont3=\tenex \scriptscriptfont3=\tenex%
  \def\it{\fam\itfam\eightit}\textfont\itfam=\eightit%
  \def\sl{\fam\slfam\eightsl}\textfont\slfam=\eightsl%
  \def\bf{\fam\bffam\eightbf}\textfont\bffam=\eightbf \scriptfont\bffam=\sixbf%
  \scriptscriptfont\bffam=\fivebf%
  \normalbaselineskip=9pt%
  \let\sc=\sixrm%
  \let\big=\eightbig%
  \setbox\strutbox=\hbox{\vrule height7pt depth2pt width\z@}%
  \normalbaselines\rm}

 % use after $ in ninepoint sections
\def\tenbig#1{{\hbox{$\left#1\vbox to8.5pt{}\right.\n@space$}}}
\def\ninebig#1{{\hbox{$\textfont0=\tenrm\textfont2=\tensy
  \left#1\vbox to7.25pt{}\right.\n@space$}}}
\def\eightbig#1{{\hbox{$\textfont0=\ninerm\textfont2=\ninesy
  \left#1\vbox to6.5pt{}\right.\n@space$}}}

% Page layout
%\newinsert\footins
\def\footnote#1{\edef\@sf{\spacefactor\the\spacefactor}#1\@sf
      \insert\footins\bgroup\eightpoint
      \interlinepenalty100 \let\par=\endgraf
        \leftskip=\z@skip \rightskip=\z@skip
        \splittopskip=10pt plus 1pt minus 1pt \floatingpenalty=20000
        \smallskip\item{#1}\bgroup\strut\aftergroup\@foot\let\next}
\skip\footins=12pt plus 2pt minus 4pt % space added when footnote is present
%\count\footins=1000 % footnote magnification factor (1 to 1)
\dimen\footins=30pc % maximum footnotes per page

\newinsert\margin
\dimen\margin=\maxdimen
%\count\margin=0 \skip\margin=0pt % marginal inserts take up no space

\loadtruetwelvepoint % At FNAL...
\loadtrueseventeenpoint
\catcode`\@=\active
\catcode`@=12  % No longer.
\catcode`\"=\active

% \use\cs
% puts in the expansion of `\cs' if it's defined, the literal "\cs" otherwise.
\def\eatOne#1{}
\def\ifundef#1{\expandafter\ifx%
\csname\expandafter\eatOne\string#1\endcsname\relax}
\def\notTrue{\iffalse}\def\isTrue{\iftrue}
\def\ifdef#1{{\ifundef#1%
\aftergroup\notTrue\else\aftergroup\isTrue\fi}}
\def\use#1{\ifundef#1\linemessage{Warning: \string#1 is undefined.}%
{\tt \string#1}\else#1\fi}

%     \ref\label{text}
% generates a number, assigns it to \label, generates an entry.
% To list the refs on a separate page,  \listrefs
% \nref does the same without generating any text at the reference
% point

\global\newcount\refno \global\refno=1
\newwrite\rfile
\newlinechar=`\^^J
\def\ref#1#2{\the\refno\nref#1{#2}}
\def\nref#1#2{\xdef#1{\the\refno}%
\ifnum\refno=1\immediate\openout\rfile=refs.tmp\fi%
\immediate\write\rfile{\noexpand\item{[\noexpand#1]\ }#2.}%
\global\advance\refno by1}
% To start a long reference...
\def\lref#1#2{\the\refno\xdef#1{\the\refno}%
\ifnum\refno=1\immediate\openout\rfile=refs.tmp\fi%
\immediate\write\rfile{\noexpand\item{[\noexpand#1]\ }#2\semi}%
\global\advance\refno by1}
% To continue a long reference...
\def\cref#1{\immediate\write\rfile{#1\semi}}
% To end a long reference...

\def\semi{;\hfil\noexpand\break}

\def\inputAuxIfPresent#1{\immediate\openin1=#1
\ifeof1\message{No file \auxfileName; I'll create one.
}\else\closein1\relax\input\auxfileName\fi%
}
% For references, some journal names

% An .aux file --- for forward references...
\newif\ifWritingAuxFile
\newwrite\auxfile
\def\SetUpAuxFile{%
\xdef\auxfileName{\jobname.aux}%
% Read it in if it exists
\inputAuxIfPresent{\auxfileName}%
% Now write a new one.
\WritingAuxFiletrue%
\immediate\openout\auxfile=\auxfileName}

% Some generally useful notation

% Warn about changed counters...
\def\bye{\par\vfill\supereject%
\ifAnyCounterChanged\linemessage{
Some counters have changed.  Re-run tex to fix them up.}\fi%
\end}

%%%%%%%%%%%%%%%%%%%%%%%%%%%%%%%%%%%%%%%%%%%%%%%%%%%%%%%%%%%%%%%%%%%

\SetUpAuxFile

\def\c{\mskip 1mu\cdot\mskip 1mu }
\def\Tr{\mathop{\rm Tr}\nolimits}

\def\sign{\mathop{\rm sign}\nolimits}

\def\eps{\epsilon}

\def\pol{\varepsilon}

\def\dl^#1_#2{\delta^{#1}{}_{#2}}

\def\Gbd{\dot G_B}
\def\Gbdd{\ddot G_B}

\def\Ord{{\cal O}}

\catcode`@=11  % Make @ letter.
\def\meqalign#1{\,\vcenter{\openup1\jot\m@th
   \ialign{\strut\hfil$\displaystyle{##}$ && $\displaystyle{{}##}$\hfil
             \crcr#1\crcr}}\,}
\catcode`@=12  % No longer.

\def\n#1#2{\nu_{#1#2}}

% counters

% General parameters
\baselineskip 15pt
\overfullrule 0.5pt

%%%%%%%%%%%%%%%%%%%%%%%%%%%%%%%%%%%%%%%%%%%%%%%%%%%%%%%%%%%%%%%%%%%%%%%%%%

%\input spinordef

\def\Tr{\mathop{\rm Tr}\nolimits}

\def\pol{\varepsilon}

\def\c{\,\cdot\,}

\def\spa#1.#2{\left\langle#1\,#2\right\rangle}
\def\spb#1.#2{\left[#1\,#2\right]}
\def\lor#1.#2{\left(#1\,#2\right)}
\def\sand#1.#2.#3{%
\left\langle\smash{#1}{\vphantom1}^{-}\right|{#2}%
\left|\smash{#3}{\vphantom1}^{-}\right\rangle}
\def\sandp#1.#2.#3{%
\left\langle\smash{#1}{\vphantom1}^{-}\right|{#2}%
\left|\smash{#3}{\vphantom1}^{+}\right\rangle}
\def\sandpp#1.#2.#3{%
\left\langle\smash{#1}{\vphantom1}^{+}\right|{#2}%
\left|\smash{#3}{\vphantom1}^{+}\right\rangle}
\catcode`@=11  % Make @ letter.
\def\meqalign#1{\,\vcenter{\openup1\jot\m@th
   \ialign{\strut\hfil$\displaystyle{##}$ && $\displaystyle{{}##}$\hfil
             \crcr#1\crcr}}\,}
\catcode`@=12  % No longer.

\loadfourteenpoint

%%%%%%%%%%%%%%%%%%%%%%%%%%%%%%%%%%%%%%%%%%%%%%%%%%%%%%%%%%%%%%%%%%%%%%%%%

\def\ref#1#2{\nref#1{#2}}
\overfullrule 0pt
\hfuzz 40pt
\hsize 6. truein
\vsize 8.5 truein

\def\eps{\epsilon}
\def\pol{\varepsilon}

\def\c{\,\cdot\,}

\def\eps{\epsilon}

\def\x#1#2{x_{#1 #2}}

\chardef\hyphen=45

%%%%%%%%%%%%%%%%%%%%%%%%%%%%%%%%%%%%%%%%%%%%%%%%%%%%%%%%%%%%%%%%%%%%%%%%%

% The references:

\ref\Long{
Z. Bern and D.A.\ Kosower, Phys.\ Rev.\ Lett.\ 66:1669 (1991);
Nucl.\ Phys.\ B379:451 (1992)\semi
Z. Bern and D.A.\ Kosower, in {\it Proceedings of the PASCOS-91
Symposium}, eds.\ P. Nath and S. Reucroft}

\ref\Bosonic{Z. Bern, Phys.\ Lett.\ 296B:85 (1992)}

\ref\AmplLet{Z. Bern, L. Dixon and D.A. Kosower, Phys.\ Rev. Lett.\
70:2677 (1993)}

\ref\Tasi{Z. Bern, UCLA/93/TEP/5, hep-ph/9304249, proceedings of TASI 1992}

\ref\TreeLevel{F.A.\ Berends and W.T.\ Giele,
Nucl.\ Phys.\ B294:700 (1987)\semi
M.\ Mangano and S.J.\ Parke, Nucl.\ Phys.
B299:673 (1988)\semi
M.\ Mangano, S. Parke and Z.\ Xu, Nucl.\ Phys.\ B298:653 (1988)\semi
M.\ Mangano, Nucl.\ Phys.\ B309:461 (1988)}

\ref\Recursive{F.A. Berends and W.T. Giele, Nucl.\ Phys.\ B306:759 (1988)\semi
D.A. Kosower, Nucl.\ Phys.\ B335:23 (1990)}

\ref\SpinorHelicity{%
F.\ A.\ Berends, R.\ Kleiss, P.\ De Causmaecker, R.\ Gastmans and T.\ T.\ Wu,
        Phys.\ Lett.\ 103B:124 (1981)\semi
P.\ De Causmaeker, R.\ Gastmans,  W.\ Troost and  T.\ T.\ Wu,
Nucl. Phys. B206:53 (1982)\semi
R.\ Kleiss and W.\ J.\ Stirling,
   Nucl.\ Phys.\ B262:235 (1985)\semi
   J.\ F.\ Gunion and Z.\ Kunszt, Phys.\ Lett.\ 161B:333 (1985)\semi
 R.\ Gastmans and T.T.\ Wu,
{\it The Ubiquitous Photon: Helicity Method for QED and QCD} (Clarendon Press)
(1990) \semi
Z. Xu, D.-H.\ Zhang and L. Chang, Nucl.\ Phys.\ B291:392 (1987)}

\ref\ManganoReview{M. Mangano and S.J. Parke, Phys.\ Rep.\ 200:301 (1991)}

\ref\Mapping{Z. Bern and D.C. Dunbar,  Nucl.\ Phys.\ B379:562 (1992)}

\ref\Gravity{Z. Bern, D.C. Dunbar and T. Shimada, Phys.\ Lett.\ B312:277
(1993)}

\ref\Unpublished{Z. Bern, L. Dixon and D. Kosower, unpublished}

\ref\Kunszt{Z. Kunszt, A. Signer and Z. Trocsanyi, preprint  ETH-TH/93-11,
 hep-ph/9305239, to appear in Nucl.\ Phys.\ B}

\ref\Matt{M.J.\ Strassler,  Nucl.\ Phys.\ B385:145 (1992)\semi
M.G.\ Schmidt and C. Schubert, preprint HD-THEP-93-24, hep-th/9309055}

\ref\Fermion{Z. Bern, L. Dixon and D.A. Kosower, in preparation}

\ref\Weak{Z. Bern and A.G.\ Morgan, preprint UCLA/93/TEP/36, DTP/93/80}

\ref\PreviousWeak{M.\ Baillargeon and F. Boudjema, Phys.\ Lett.\ B272:158
(1991)\semi
X.Y. Pham, Phys.\ Lett.\ B272:373 (1991)\semi
F.-X.\ Dong, X.-D. Jiang and X.-J. Zhou, Beijing preprint BIHEP-TH-92-32\semi
E.W.N.\ Glover and A.G.\ Morgan, DPT/93/4}

\ref\ChanPaton{J.E.\ Paton and H.M.\ Chan, Nucl.\ Phys.\ B10:516 (1969)}

\ref\GSB{M.B.\ Green, J.H.\ Schwarz and L. Brink, Nucl.\ Phys.\
B198:472 (1982)}

\ref\Scherk{J. Scherk, Nucl.\ Phys.\ {B31} (1971) 222\semi
A. Neveu and J. Scherk, Nucl.\ Phys.\ {B36} (1972) 155\semi
J.\ Minahan, Nucl.\ Phys.\ B298:36 (1988)}

\ref\Heterotic{D.J. Gross, J.A. Harvey, E. Martinec and R. Rohm,
Nucl.\ Phys.\ B256:253 (1985); Nucl.\ Phys.\ B267:75 (1986)}

\ref\Beta{Z.\ Bern and D.A.\ Kosower, Phys.\ Rev.\ D38:1888 (1988)\semi
Z.\ Bern and D. A.\ Kosower, in proceedings, {\it Perspectives in String
Theory}, eds. P.\ Di Vecchia and J. L.\ Petersen, Copenhagen 1987}

\ref\KLT{
L.\ Dixon, J.\ Harvey, C.\ Vafa and E.\ Witten, Nucl.\ Phys.\
{B261}:678 (1985), {B274}:285 (1986)\semi
K.S.\ Narain, Phys.\ Lett.\ {169B}:41 (1986)\semi
W.\ Lerche, D.\ Lust and A.N.\ Schellekens, Nucl.\ Phys.\
{B287}:477 (1987)\semi
H.\ Kawai, D.C.\ Lewellen and S.-H.H.\ Tye, Nucl.\ Phys.\
B288:1 (1987)\semi
K.S.\ Narain, M.H.\ Sarmadi and C.\ Vafa,
Nucl.\ Phys.\ {B288}:551 (1987)\semi
I. Antoniadis, C.P.\ Bachas and C.\ Kounnas, Nucl.\ Phys.\
{B289}:87 (1987)}

\ref\SchwarzReview{J.H.\ Schwarz, Phys. Reports 89:223 (1982)}

\ref\typeII{L.\ Dixon, V.\ Kaplunovsky and C. Vafa, Nucl.\ Phys.\
B294:43 (1987)\semi
H.\ Kawai, D.C.\ Lewellen and S.-H. H.\ Tye, Phys.\ Lett.\ 191B:63 (1987)}

\ref\OpenString{Z.\ Bern and D.C.\ Dunbar,  Phys.\ Rev.\ Lett.\ 64:827 (1990);
          Phys.\ Lett.\ 242B:175 (1990)}

\ref\Tseytlin{R.R.\ Metsaev and A.A.\ Tseytlin, Nucl.\ Phys.\ B298:109 (1988)}

\ref\Polyakov{A.M.\ Polyakov, Phys.\ Lett.\ 103B:207 (1981); 103B:211 (1981)}

\ref\GSW{M.B.\ Green, J.H.\ Schwarz,
and E.\ Witten, {\it Superstring Theory} (Cambridge University
Press) (1987)}

\ref\Lovelace{C. Lovelace, Phys.\ Lett.\ 34B:500 (1971)}

\ref\Color{Z. Bern and D.A.\ Kosower, Nucl.\ Phys.\ B362:389 (1991)}

\ref\Background{G. 't Hooft,
{\it in} Acta Universitatis Wratislavensis no.\
38, 12th Winter School of Theoretical Physics in Karpacz; {\it
Functional and Probabilistic Methods in Quantum Field Theory},
Vol. 1 (1975)\semi
B.S.\ DeWitt, {\it in} Quantum gravity II, eds. C. Isham, R.\ Penrose and
D.\ Sciama (Oxford, 1981)\semi
L.F.\ Abbott, Nucl.\ Phys.\ B185:189 (1981)\semi
L.F. Abbott, M.T. Grisaru and R.K. Schaefer,
Nucl.\ Phys.\ B229:372 (1983)}

\ref\GN{J.L.\ Gervais and A. Neveu, Nucl.\ Phys.\ B46:381 (1972)}

\ref\Lam{J.D. Bjorken, Stanford Ph.D. thesis (1958)\semi
J.D. Bjorken and S.D. Drell, {\it Relativistic Quantum Fields}
(McGraw-Hill, 1965)\semi
J. Mathews, Phys.\ Rev. 113:381 (1959)\semi
S. Coleman and R. Norton, Nuovo Cimento 38:438 (1965)\semi
C.S. Lam and J.P. Lebrun, Nuovo Cimento 59A:397 (1969\semi
C.S. Lam,  Nucl.\ Phys.\ B397:143 (1993)}

\ref\Integrals{Z. Bern, L. Dixon and D.A. Kosower,
Phys.\ Lett.\ B302:299,1993; SLAC--PUB--5947, Nucl.\ Phys.\ B, to appear\semi
R.K.\ Ellis, W. Giele and E. Yehudai, in progress}

\ref\Ellis{R.K.\ Ellis and J.C.\ Sexton, Nucl.\ Phys.\ B269:445 (1986)}

\ref\BerendsGrav{F.A. Berends, W.T.\ Giele and H. Kuijf,
Phys. Lett.\ {211B}:91 (1988)}

\ref\Susy{
M.T.\ Grisaru, H.N.\ Pendleton and P.\ van Nieuwenhuizen,
Phys. Rev. {D15}:996 (1977)\semi
M.T. Grisaru and H.N. Pendleton, Nucl.\ Phys.\ B124:81 (1977)\semi
S.J. Parke and T. Taylor, Phys.\ Lett.\ B157:81 (1985)\semi
Z. Kunszt, Nucl.\ Phys.\ B271:333 (1986)\semi
M.L.\ Mangano and S.J. Parke, Phys.\ Rep.\ {200}:301 (1991)}

\ref\AllN{Z. Bern, G. Chalmers, L. Dixon, D. Kosower, in preparation}

\ref\ParkeTaylor{S.J. Parke and T.R. Taylor, Phys.\ Rev.\ Lett.\ 56:2459,1986}

\ref\Mahlon{G.D.\ Mahlon, preprint Fermilab-Pub-93/327-T, hep-ph/9311213}

\ref\Roland{K. Roland, Phys.\ Lett.\ B289:148 (1992);
preprint SISSA/ISAS 131-93-EP\semi
G. Cristofano, R. Marotta and K. Roland,
Nucl.\ Phys.\ B392:345 (1993)}

%%%%%%%%%%%%%%%%%%%%%%%%%%%%%%%%%%%%%%%%%%%%%%%%%%%%%%%%%%%%%%%%%%%%%%%%%%%%%

\noindent
hep-ph/9311026 \hfill Saclay/SPhT--T93/116

\hfill SLAC-PUB-6388

\hfill UCLA/93/TEP/40

\hskip .3 cm

\baselineskip 12 pt
\centerline{{\bf NEW QCD RESULTS FROM STRING THEORY}%
\footnote{${}^*$}%
{Talk presented by Z.B. at Strings 1993, May 24-29, Berkeley CA} }

\vskip .3 cm
\centerline{\ninerm ZVI BERN}
\baselineskip=13pt
\centerline{\nineit Department of Physics, UCLA, Los Angeles, CA 90024}
\vglue 0.3cm

\centerline{\ninerm LANCE DIXON}
\centerline{\nineit Stanford Linear Accelerator Center, Stanford University,
Stanford, CA 94309}

\vglue 0.2cm
\centerline{\ninerm and}
\vglue 0.2cm
\centerline{\ninerm DAVID A. KOSOWER}
\baselineskip12truept
\centerline{\nineit Service de Physique Th\'eorique de Saclay,
 Centre d'Etudes de Saclay}
\centerline{\nineit F-91191 Gif-sur-Yvette cedex, France}

\vglue 0.7cm
\centerline{\tenrm ABSTRACT}
\vglue 0.3cm
{\rightskip=3pc
\leftskip=3pc
\tenrm\baselineskip=12pt%\parindent=1pc
\noindent
We discuss new results in QCD obtained with string-based methods.
These methods were originally derived from superstring
theory and are significantly more efficient than conventional Feynman
rules.  This technology was a key ingredient in the first
calculation of the one-loop five-gluon amplitude.  We also present
a conjecture for a particular one-loop
helicity amplitude with an arbitrary number
of external gluons.
}

\baselineskip 12 pt

\vglue 0.5cm

\noindent
{\bf 1. Introduction.}

\vskip .1 cm

Calculations beyond the leading order in quantum chromodynamics are
important to refining our understanding of known physics in
present-day and future collider experiments.
The discovery of new physics relies to a large
extent on the subtraction of known physics from the data.  In particular, QCD
loop corrections are important but are in general quite formidable to
calculate.  Intermediate expressions can be many thousands of times
larger in size than final expressions.  This explosion of terms has
been a major obstacle in performing computations required by
experiment.  Here we discuss new techniques based on string theory
which bypass much of the algebra associated with one-loop Feynman
diagram computations in gauge theories [\use\Long --\use\Tasi].
With the new string-based techniques the one-loop five-gluon
amplitudes have been computed yielding a compact form [\use\AmplLet].
These amplitudes have not been obtained with traditional techniques.

Recent years have seen substantial progress in improving the situation
in tree-level calculations [\use\TreeLevel-\use\ManganoReview].
Tree-level matrix elements have been essential for checks of QCD
processes and for estimates of QCD backgrounds to new physics
searches.  Here we discuss the one-loop corrections to the amplitudes.
The next-to-leading order corrections are important because leading tree-level
calculations miss essential physics. Two problems are that the
tree-level results exhibit a strong renormalization scale dependence
(which does not make physical sense) and that the cone angle
dependence is incorrect.  The next-to-leading order corrections to a
large extent remedy this situation.

The one-loop method discussed here was originally derived
from string theory.  Although based on string theory it has been
summarized in terms of simple rules which require no knowledge of
string theory [\use\Long]; the structure of these rules can also
be understood from conventional field theory through particular
gauge choices and organizations of the amplitude [\use\Mapping].

Using methods developed through string theory we have obtained results
which have previously not been obtained through conventional means.
Besides the five-gluon amplitudes [\use\AmplLet], a first calculation
of a four-point one-loop gravity amplitude has been performed
[\use\Gravity].  We have used the explicit
five-gluon results to jump-start a conjecture for a particular
one-loop helicity amplitude but with arbitrary numbers of external
legs [\use\AllN].

Using direct string methods we have also calculated
the one-loop four-point helicity amplitudes with two external quarks
[\use\Unpublished], which agree with the calculation of ref.~[\use\Kunszt].

Is string theory `required' for field theory calculations?  To develop
and extend the methods string theory has been crucial.  To actually
evaluate amplitudes there is no need to turn to string theory. The
main role of string theory is to provide a principle for discovering
compact representations for field theory amplitudes.  In particular,
given the string-based rules for the one-loop $n$-gluon amplitudes and
the understanding of these rules in field theory
[\use\Mapping,\use\Matt], there does not appear to be a clear way to
extend the rules to multi-loops, or to gravity, without referring back to
string theory to at least some extent.  It is, however, possible to
formulate a conventional field theory framework for obtaining much of
the efficiency of the string-based method by working backwards from
the string-based rules.  These field theory ideas can then be applied
more generally to gauge theory amplitude calculations which involve
non-abelian vertices [\use\Mapping].  For example, some of the ideas
obtained from string theory were used to aid the calculation of
one-loop five-point amplitudes with external fermions and gluons
[\use\Fermion].  Another example is the application of these ideas
[\use\Weak] to weak interaction processes such as $Z\rightarrow 3
\gamma$ [\use\PreviousWeak].

\vskip .2 cm
\noindent{\bf 2. Difficulty of Loop Computations}
\vskip .1 cm

An underlying cause of the complexity of QCD calculations is that the
non-abelian vertices are relatively complicated. Since the vertices
each contain six terms, one encounters a rapidly growing number of
terms as one sews together vertices with propagators to form Feynman
diagrams.  Furthermore, the integrals associated with larger numbers
of legs become increasingly complicated.

As a simple example consider the pentagon diagram one
would encounter in a brute force five-point computation.  A naive count
of the number of terms gives about $6^5$ terms.  (This count
is reduced by the use of on-shell conditions but
increased since each internal momentum turns into a sum of external
momenta.)  Each term is associated with an integral which
evaluates to an expression on the order of a page in length.
This means that one is faced with about $10^4$ pages of algebra
for this single diagram.   As bad as this situation might seem, it is
actually much worse because of the structure of the results.
After evaluating the integrals and summing over diagrams
one obtains expressions of the form
$$
{N_1 \over D_1} + {N_2\over D_2} + \cdots
\anoneqn
$$
where the $N_i$ and $D_i$ are the numerators and denominators one
encounters when performing the integrals.  In general the denominators
contain spurious singularities which cancel only after putting large
numbers of terms on a common denominator; this unfortunately causes an
explosion of terms in the numerators.

The basic observation for being able to improve on conventional computations
is that Feynman diagram computations always involve large cancellations
amongst the various terms.  Anyone who has done a
Feynman diagram computation has undoubtedly asked themselves why
vasts amounts of algebra are required when answers tend to be
quite small.  A nice example of this is the one-loop
four-gluon helicity amplitude
$$
A_{4;1} (1^-, 2^+, 3^+, 4^+) = - \Bigl( 1 + {n_s\over N} -
{n_f\over N} \Bigr)
 {i \over 48 \pi^2} {\spb2.4{}^2 u \over \spb1.2 \spa2.3 \spa3.4 \spb4.1 }
\anoneqn
$$
where $n_s$ and $n_{\! f}$ are the number of
massless complex scalars and Weyl fermions in the fundamental
representation of an $SU(N)$ theory.
The plus and minus signs associated with each leg denote the
helicity and $u= 2k_1\c k_3$ is a Mandelstam variable.
The various brackets are spinor inner products defined by
[\use\SpinorHelicity]
$$
\spa{j}.l = \langle {k_j}^{-} | {k_l}^{+}\rangle , \hskip 1.5 cm
\spb{j}.l = \langle {k_j}^{+} | {k_l}^{-}\rangle
$$
where the $|{k_l}^{\pm}\rangle$ are Weyl spinors.
The spinor inner products
are a convenient way to represent helicity amplitudes.
(This amplitude has been decomposed using the
one-loop Chan-Paton color decomposition [\use\ChanPaton].)
Although this expression fits on a
line, a brute force computation performed in the conventional way
would start with expressions containing about $10^4$ terms. Clearly
there is considerable room for improving Feynman diagram computations
at one loop.  The claim is that string theory coupled with other
ideas such as spinor helicity methods provides a way for
doing this.

\break
%\vskip .2 cm
\noindent{\bf 3. Why String Theory is Helpful.}
\vskip .1 cm

It does not take much string theory for one to realize that string
theory can be helpful in gauge theory computations.  String theory
satisfies a number of properties which indicates its usefulness in
gauge theory computations.

\item{1)} String amplitudes are compact.  At each order of perturbation
theory there is only a single diagram; this provides for a compact
organization, which might lead one to suspect (at least at a hand-waving
level) that string theory might be useful for gauge theory
calculations.

\item{2)} In string theory one can switch between fermions and bosons
in the loop by changing the world-sheet boundary conditions.  This
means that whether bosons or fermions circulate in the loop, the same
basic string equations describe the two situations.  This leads one to
suspect that it should be possible to use information from a
calculation with fermions in the loop to aid calclations with bosons
in the loop. This may be contrasted to the usual Feynman rules one
finds in textbooks, where the fermion and gluon Feynman rules do not
resemble each other.

\item{3)} The $N=4$ superstring amplitude integrands are
simple [\use\GSB]. In particular for the four-point amplitude
the kinematic expression factors out of the string integrand.
In the field theory limit this implies that the
various contributions to a gluon amplitude $A$ satisfy
$$
 A^{\rm gluon} + 4 A^{\rm fermion} + 3 A^{\rm scalar} = \hbox{simple}
\anoneqn
$$
where the particle labels refer to the
particles circulating in the loop.
(Note that the massless spectrum of $N=4$ super-Yang-Mills
is one gluon, four Weyl fermions and three
complex scalars.)  This relation then imples that not
only are the integrands of the various contributions related via the
previous point, but that they in fact satisfy a simple linear relation.
These relations can be used to prevent duplication of effort when
calculating the various contributions to a process.

\item{4)} In string theory (closed string) $\sim$ (open string)$^2$.
Since closed strings contain gravity and open strings contain
Yang-Mills, one might expect that (gravity) $\sim$ (Yang-Mills)$^2$.
This can be made precise and be used to turn string theory into an
extremely efficient computational tool for gravity [\use\Gravity].

\item{5)} In string theory the loop momentum is implicitly integrated
out.  This is useful because helicity techniques [\use\SpinorHelicity]
are most efficient after loop momentum is integrated out.

The strategy is to build a string theory with a spectrum relevant
for the field theory of interest, and then to take the infinite
string tension limit in order to extract the organizational efficiency
of the string in the field theory limit.  Once the essential reorganization
has been summarized in terms of a set of field theory rules or ideas
there is no need to turn to string theory for every new calculation.

\break
%\vskip .2 cm
\noindent
{\bf 4. String Theory and Extraction of Field Theory Limit}
\vskip .1 cm

The infinite tension limit of a string theory is a field theory
[\use\Scherk,\use\GSB].  In
order to use string theory as a computational tool, control of the
massless matter content of the string theory is required, because
colored massless matter particles can run around the loops.  It is
possible to build consistent heterotic string theories
[\use\Heterotic] whose infinite-tension limit is a non-abelian gauge
theory where one of the factors is an $SU(N)$ with no matter fields
[\use\Beta].  The technology needed for such a construction is the one
used to construct four-dimensional string models [\use\KLT]; the
formulation of Kawai, Lewellen and Tye is particularly simple,
although any of the other formulations can be used depending on one's
taste.  In the original derivation of the string-based rules
[\use\Long], it was essential to use a consistent string
in order to prevent extraneous problems from entering.  Without full
consistency there would be no guarantee that the final results
obtained would be correct.  A heterotic string was used in the
original derivation of the string-based rules because bosonic
strings always contain unwanted massless scalars and tachyons, while
four-dimensional type~II [\use\SchwarzReview,\use\typeII]
and type~I [\use\OpenString] superstrings do not have a rich enough
variety of fully consistent models.

Bosonic string constructions are generally much simpler than super or
heterotic string constructions so for pedagogical purposes that is
what will be discussed here.  Heterotic string constructions are
needed to provide a consistent derivation of the rules [\use\Long],
but the procedure for extracting the field theory limit is similar.
The open bosonic string discussed here is identical to the one used by
Metsaev and Tseytlin [\use\Tseytlin] to obtain the Yang-Mills
$\beta$-function from string theory.  This string is given by a naive
truncation of an oriented open bosonic string to four-dimensions.  In
this way all massless colored scalars arising from the dimensional
compactification are simply thrown away.  This string is inconsistent
as a fundamental string theory because of the naive truncation of the
spectrum.  Another technicality is that the string does contain a
tachyon, which might be worrisome; however, one can handle this with
the prescription that exponentially large terms due to the tachyon
should be dropped in the same way that exponentially small terms from
the higher mass states are dropped.  These potential inconsistencies
of the bosonic string are irrelevant in the field-theory limit, as can
be verified explicitly using an independent calculation with the
fully-consistent heterotic-string formalism.  What is important here
is the basic structure that emerges from string theory without facing
the full technicalities of heterotic string constructions.

In general, an amplitude in string theory is evaluated by performing
the Polyakov surface integral [\use\Polyakov]
$$
A_n  \sim \int DX \exp\Bigl[ {1\over \alpha'}
\int d^2 \nu \; \partial_\alpha X^\mu \partial_\alpha X_\mu \Bigr]
V_1 V_2 \cdots V_n
\anoneqn
$$
where the $V_i \sim \pol_i \c \partial X e^{ik\cdot X}$ are the vertex
operators for external gluons with polarizations $\pol$.
At one-loop this path integral is
performed on a world-sheet annulus.
Since the world-sheet bosons are free,  Wick's theorem can be used
to evaluate the string $n$-gluon amplitude in terms of the two-point
correlation on the annulus
$$
\eqalign{
\langle X_{\mu}(\nu_1) X^{\nu}(\nu_2) \rangle
& = \delta_\mu{}^\nu G_B(\n 12) = - \delta_\mu{}^\nu
\Bigl[ \log\ |{2\sinh(\n 12) } | - { (\n12)^2 \over\tau }
- 4 q \sinh^2 (\n12) \Bigr] \cr
\null & \hskip 1cm
 + \Ord(q^2) \cr }
\anoneqn
$$
where $\tau=-\log(q)/2$ is the real modular parameter of the annulus,
$\nu_i$ represents the location of the vertex operator on the annulus
and $\n ij = \nu_i - \nu_j$.
(These parameters are $\pi/i$ times the conventional one in
refs.~[\use\SchwarzReview,\use\GSW].)
As discussed in ref.~[\use\Long], in the field theory limit
these parameters are proportional to
sums of Schwinger proper time parameters.
A repeated application of Wick's theorem to evaluate the Polyakov integral
yields the string partial amplitude associated with the color trace
$\Tr(T^{a_1} T^{a_2} \cdots T^{a_n})$
$$
\eqalign{
A_{n;1} = & i {(4\pi)^{\eps/2} \over 16 \pi^2}
(\sqrt{2})^n (\alpha')^{n/2 - 2}
\int_0^\infty d\tau  \int \prod_{i=1}^{n-1} d \nu_i\theta(\nu_i-\nu_{i+1})\;
{\tau}^{-2+\eps/2} Z \cr
& \null \hskip .5 cm \times
\prod_{i<j}^n \exp \biggl\{
\alpha' k_i\c k_j G_B(\n ij)  +
\sqrt{\alpha'}(k_i\c\pol_j - k_j\c\pol_i ) \, \Gbd(\n ij) \cr
\null & \hskip 3 cm
- \pol_i\c\pol_j\, \Gbdd(\n ij) \biggr\}
\biggr|_{\rm multi-linear}
\cr}
\eqn\StringAmplitude
$$
where
$$
\Gbd(\nu) = {1\over 2} {\partial \over \partial \nu} G_B(\nu) \; ,
\hskip 2 cm
\Gbdd(\nu) = {1\over 4} {\partial^2 \over \partial \nu^2} G_B(\nu)
\anoneqn
$$
and $\nu_n$ is fixed at $\tau$.  The `multi-linear' signifies that
after expanding the exponential only terms which are linear in all $n$
polarizations vectors are to be kept. The
string oscillator contributions to the partition function are
$$
Z =  q^{-1}
\prod_{n=1}^\infty(1 - q^n)^{-2(1-\delta_R\eps/2)} \; .
\eqn\StringPartFunc
$$
Full consistency of the string demands that the dimension
$D=26$ [\use\Lovelace], but for the purposes of obtaining
field theory amplitudes $D=4-\eps$ where $\eps$ is the dimensional
regularization parameter necessary to handle infrared divergences; the
regularization parameter $\delta_R$, included in the string partition
function, determines the precise form of the regularization
[\use\Long]. In order to obtain a sensible field
theory limit, the leading $q^{-1}$ has been maintained by hand independent
of the number of dimensions.  (A fully consistent heterotic string
such as the one used in ref.~[\use\Long]
does not require any adjustments, such as this one.)
The field theory limit of the amplitude (\use\StringAmplitude) yields the pure
Yang-Mills contributions to the amplitude including Faddeev-Popov
ghosts.  The conventions have been adjusted so that in the field
theory limit the number of $\pi$'s and 2's which need to be shuffled
around are minimized and so that these normalizations agree with the
ones used in the heterotic string analysis of refs.~[\use\Long].

Partial amplitudes associated with two color traces
are a bit different since the string vertex operators
are located on both boundaries of the annulus; examples can be
found in chapter 8 of ref.~[\use\GSW].

In order to take the infinite string tension limit $\alpha'
\rightarrow 0$ of the string amplitude (\use\StringAmplitude), it is
convenient to first integrate by parts on the string world-sheet in
order to remove all $\Gbdd$ from the kinematic factor
[\use\Long].  (The analysis of the field theory limit can
also be performed without the integration-by-parts step
so it should not be taken as an essential ingredient to
the string-based method.)  In open string theory there are potential
boundary terms, but these can be removed by an appropriate analytic
continuation in external momenta since all the boundary terms contain
a factor of $|\nu_i -
\nu_j|^{-n - \alpha' k_i \c k_j} |_{\nu_i\rightarrow\nu_j} = 0$.
(One technicality is that the periodicity on the annulus under $\nu
\rightarrow \nu+\tau$ must be used to remove some of the surface
terms.)
In appendix B of ref.~[\use\Color] it was proven that
all $\Gbdd$'s can always be eliminated from the kinematic function,
by appropriate integration by parts.

In the field theory limit, the contributions to an integrated-by-parts
one-loop amplitude can be classified in terms of tree and loop parts.
The tree parts are obtained by first extracting the massless poles in
the $S$-matrix before taking the field theory limit of the loop.
Examples of these kinematic poles are found in the regions where
$\nu_i \rightarrow \nu_j$ and are of the form
$$
\int d \nu_i {1\over \nu_{ij}^{1+ \alpha' k_i \c k_j}} \longrightarrow
- {1 \over \alpha' k_i \c k_j}  \hskip 1.5 cm (\alpha' \rightarrow 0) \; .
\anoneqn
$$
In general, the kinematic poles extracted in this way correspond to the poles
of a scalar $\phi^3$ diagram.

After kinematic poles have been extracted, the field theory limit of the
loop is needed.
This is obtained by taking $\tau, |\nu| \rightarrow \infty$ which corresponds
to squeezing the annulus down to a field theory loop.
The values of the Green functions in this limit are
$$
\eqalign{
\exp(G_B(\nu)) &\rightarrow
\exp\Bigl({\nu^2\over\tau}-|\nu| \Bigr)
\times \hbox{constant} \cr
\Gbd(\nu) & \rightarrow {\nu \over \tau} - \sign(\nu) (
{\textstyle{1\over 2}}+ e^{-2|\nu|} - q e^{2|\nu|}  ) \; .  \cr}
\anoneqn
$$
The exponentiated bosonic Green function was not expanded
beyond $\Ord(q^0)$; after carrying out the integration by parts procedure
the higher order terms do not contribute
since they carry too many explicit powers of $\alpha'$.  For
$\Gbd$, terms through $\Ord(q)$ should be kept due to the presence
of the overall $q^{-1}$ in the string amplitude (\StringAmplitude).

In the field theory limit two types of loop contributions are obtained
depending on whether a power of $q$ is extracted from the string
partition function or from the Green functions.  For the former
contribution one simply keeps the leading order contributions from the
bosonic Green functions. This type of contribution is described by the
bosonic zero-mode [\use\GSW] or loop momentum integral of the string
[\use\Mapping].
A product of $\Gbd$'s contains exponentially
growing and decaying terms as well as terms which are constant.
In general, when terms proportional to $q= e^{-2 \tau}$ are extracted
from a product of $\Gbd$ in order to cancel the overall $q^{-1}$,
a factor of the form
$$
\exp \Bigl[\Bigl( |x_k - x_l| - \sum |x_i - x_j| \Bigr) \tau \Bigr]
\anoneqn
$$
is obtained where $\x_i\equiv \nu_i / \tau$.
In order to avoid exponential suppression or growth as $\alpha' \rightarrow 0$
the sum must add up to cancel within the exponential exactly.
This will happen only if each $x_i$ which appears with a positive sign
also appears with a negative sign after expressing the absolute
values in terms of the $x_i$s directly.
The correct prescription for
dealing with exponentially growing terms due to the tachyon
is to simply drop them
in the same way that exponentially decaying terms are dropped. (The
exponential growth is an artifact of the Schwinger proper
time representation of tachyonic propagators.)

The result of collecting those terms where the exponential
terms completely cancel is that only those which form a {\it cycle}
of $\Gbd$'s, defined to be a product of $\Gbd$'s
with indices arranged in the form
$$
\Gbd(\n {i_1}{i_2}) \Gbd(\n {i_2}{i_3})  \cdots \Gbd(\n {i_m} {i_1}) \; ,
\anoneqn
$$
will not vanish.  Furthermore, the cyclic ordering of the indices must
follow the same ordering of the corresponding legs in the partial
amplitude.

The analysis of the field theory limit of a superstring is quite
similar.  A superstring is essential in order to be able to include
space-time fermions into the string formalism and provides a more
consistent framework than bosonic strings.  A detailed discussion
of the field theory limit of a heterotic string has been given in
ref.~[\use\Long].

By organizing the contributions obtained in the field theory limit a
set of string-based rules for calculating gluon amplitudes can be
obtained [\use\Long,\use\Bosonic].
The string-based rules work by specifying a set of substitution
rules on  the string kinematic expression
$$
\eqalign{
{\cal K} &=
\int \prod_{i=1}^n dx_i \prod_{i<j}^n
\exp\biggl( k_i\c k_j G_B^{i,j} +
(k_i\c\pol_j - k_j\c\pol_i) \, \Gbd^{i,j}
- \pol_i\c\pol_j\, \Gbdd^{i,j} \biggr)
\biggr|_{\rm multi-linear}
\cr}
\eqn\MasterKin
$$
to obtain the contributions to various diagrams which represent the
various corners of moduli space.
This `master formula' contains all information for all diagrams and particle
types which can circulate in the loop.
Although this is the kinematic formula for a bosonic string, one can
use world-sheet supersymmetry in a superstring to relate the contributions
of the world-sheet fermions to those of the world-sheet bosons [\use\Bosonic];
in this
way the kinematic expression associated with the bosons can be used to
summarize all contributions.  Alternatively one can
specify rules which act directly on the superstring kinematic expression,
which was the original form of the rules presented in ref.~[\use\Long].
An example of a substitution rule is
$$
\Gbd^{i,j} \rightarrow {1\over 2} + \sum_{k=i+1}^j a_k
\anoneqn
$$
where the $a_k$ are ordinary Feynman parameters.
There are a variety of other substitution rules which depend on the
particular corner of string moduli space under consideration and
are described more fully in refs.~[\use\Long,\use\Tasi].

Because the contribution of any type of particle is
contained in the master formula, relationships between fermion and
boson contributions become apparent within the integrands of each
diagram.  This can be used to obtain even further simplifications;
once the fermion loop contribution to the $n$-gluon amplitude has been
computed, calculating the gluon loop contribution is relatively simple
[\use\AmplLet].

In this way one obtains a set of Feynman parameter polynomials which
are far more compact than one would obtain by traditional Feynman
diagram methods.   The Feynman parameters must then be integrated in
order to obtain the amplitudes.

\vskip .3 cm
\noindent
{\bf 5. Field Theory Interpretation}
\vskip .1 cm

Since a conventional field theory interpretation of string-based rules
has been developed [\use\Mapping], one can use string-motivated ideas
directly in a conventional field theory setting.  The following
strategy incorporates the ideas that were extracted from the mapping
between field theory and string theory and greatly improves the
calculational efficiency over traditional Feynman diagram
computations.  First background field Feynman gauge [\use\Background]
should be used in calculations where a non-abelian vertex appears in
the loop. This gauge is useful for constructing the
one-particle-irreducible diagrams, since the vertices are particularly
well suited for loop calculations. These one-particle irreducible
diagrams describe a gauge-invariant effective action.  For sewing
trees onto the one-particle irreducible loop diagrams the
Gervais-Neveu gauge [\use\GN] is a particularly efficient gauge
because of the simplicity of the three- and four-point vertices.
(Although it might seem strange that two different gauge choices are
used for the loop and tree parts of the Feynman diagrams, in the
background field method this has been justified by Abbott, Grisaru and
Schaeffer [\use\Background].)  In general, one should use color
ordered [\use\TreeLevel,\use\Color] vertices in order to minimize the
number of diagrams which must be explicitly computed.  For internal
fermions it is best to use the second order formalism described in
ref.~[\use\Mapping] because then the gauge boson and fermion
contributions are quite similar.  In particular for $N=4$ supersymmetry
there are large cancellations between the various contributions
to the loops [\use\GSB].
In this way a good
fraction of the work does not have to be duplicated for each type of
particle circulating in the loop in agreement with the string theory
expectations.

Spinor helicity methods [\use\SpinorHelicity] are also important to
help minimize the amount of required algebra.  Since spinor helicity
methods do not handle off-shell loop momentum efficiently it should be
integrated out early in the calculation to obtain a representation in
terms of Feynman parameters.  In order to minimize the number of terms
which appear, spinor helicity should be applied on a term-by-term
basis in the numerator as one integrates out the loop momentum.  An
alternative approach which implicitly and systematically integrates
out the loop momentum is the electric circuit analogy discussed in
refs.~[\use\Lam].

In this way one can attain many of the simplifications of a more
direct string approach.  Gravity does, however, provide a concrete
example where a direct string-based computation is significantly
more efficient than a computation based on the above field theory
ideas [\use\Gravity].

\vskip .2 cm
\noindent{\bf 6. Explicit Calculations}
\vskip .1 cm

Using the string-based methods discussed above we have performed a
computation of the one-loop five gluon amplitudes [\use\AmplLet].
Additional ingredients
which enter into this calculation are a simple integration method for the
pentagon parameter integrals [\use\Integrals] and improvements in the
spinor helicity method.

The finite helicity one-loop amplitudes associated with the color trace
$\Tr(T^{a_1} \cdots T^{a_5})$ are
$$
\eqalign{
 A_{5;1}&(1^+,2^+,3^+,4^+,5^+)
=\ \Bigl(1+ {n_s \over N} - {n_{\! f}\over N} \Bigr) \cr
& \hskip 1 cm \times
{i\over 48\pi^2}\,
  {  \spa1.2\spb1.2\spa2.3\spb2.3 + \spa4.5\spb4.5\spa5.1\spb5.1
   + \spa2.3\spa4.5\spb2.5\spb3.4
 \over \spa1.2 \spa2.3 \spa3.4 \spa4.5 \spa5.1 } \cr }
\eqn\FivePlus
$$
$$
\eqalign{
 A_{5;1}&(1^-,2^+,3^+,4^+,5^+)
=\ \Bigl(1+ {n_s\over N} -  {n_{\! f}\over N} \Bigr) \cr
& \hskip 1 cm \times
{ i\over 48\pi^2}\,
{1\over\spa3.4^2 }
 \biggl[-{ \spb2.5^3 \over \spb1.2\spb5.1 }
 + { \spa1.4^3\spb4.5\spa3.5 \over \spa1.2\spa2.3\spa4.5^2 }
 - { \spa1.3^3\spb3.2\spa4.2 \over \spa1.5\spa5.4\spa3.2^2 } \biggr] \ .
  \cr }
\anoneqn
$$

The infrared-divergent amplitudes (which are the ones which interfere with
the tree diagrams to produce the next-to-leading order corrections to
the cross-section) are given in ref.~[\use\AmplLet] and are more complicated.
These amplitudes have not been obtained with traditional techniques
used, for example, by Ellis and Sexton [\use\Ellis].

We have also calculated four-point matrix elements with two external
quarks by a direct string approach [\use\Unpublished].  Five-point
matrix elements have also been obtained, using field theory.
An example of one of the amplitudes is
$$
\eqalign{
&A_{5;1}(1_{\bar{q}}^-,  2_q^+ ,3^{+},4^{+},5^{+})\ =\
  {i\over16\pi^2}\Biggl[ -{1\over2}\left( 1+{1\over N^2} \right)
 { \spa1.2
    ( \spa1.2 \spb2.3 \spa3.1
    + \spa1.4 \spb4.5 \spa5.1 )
  \over \spa1.2\spa2.3\spa3.4\spa4.5\spa5.1 } \cr
\null & \hskip .2 cm
 + {1\over3} \left( 1-{n_f\over N}+{n_s\over N} \right)
 \biggl(
  { \spa1.3\spb3.4{\spa4.1}^2
   \over \spa1.2{\spa3.4}^2\spa4.5\spa5.1 }
 + { \spa1.4\spa2.4\spb4.5\spa5.1
   \over \spa1.2\spa2.3\spa3.4{\spa4.5}^2 }
 + { \spb2.3\spb2.5 \over \spb1.2 \spa3.4 \spa4.5 }
 \biggr)  \Biggr]\ .\cr }
\eqn\qgggqfinite
$$
The remaining five-point amplitudes with external
fermions are more complicated and
will be presented elsewhere [\use\Fermion].
The calculational ideas motivated by the string organization were useful
in obtaining these results.

\vskip .2 cm
\noindent
{\bf 7. String-Based Methods for Gravity}
\vskip .1 cm

Another application of the string-based technique is to gravity
[\use\Gravity]. Roughly speaking the structure of string theory implies
that
$$
(\hbox {Closed String}) \sim (\hbox{Open String})^2 \; .
\anoneqn
$$
Since closed strings contain gravity and open strings contain gauge theory
one might expect that
$$
(\hbox{Gravity}) \sim (\hbox{Yang-Mills})^2 \; .
\anoneqn
$$
This relationship can be made precise and turned into an extremely
efficient computational tool for perturbative gravity amplitudes.
At tree-level Berends, Giele and Kuijf
[\use\BerendsGrav] have made use of this relationship
to obtain tree-level gravity amplitudes
from known Yang-Mills amplitudes.
At one-loop this relationship can also be made precise [\use\Gravity];
in particular, the calculation of the one-loop
four-graviton amplitude with one minus and three plus helicities is
rather easy by making use of string-based rules modified for the case
of gravity coupled to massless matter.
The result of such a calculation is given by
$$
A(1^-, 2^+, 3^+, 4^+) = {i \kappa^4 \over (4\pi)^2} {1\over 5760}
 (N_b - N_{\!f}) {s^2 t^2 \over u^2}
 (u^2 - st ) \biggl( {[24]^2 \over [12] \langle
23\rangle \langle 34\rangle [41] } \biggr)^2
\anoneqn
$$
where $\kappa$ is the gravitational coupling,
$N_b$ is the number of physical bosonic states
and $N_{\!f}$ is the number of
fermionic states in the particular theory of gravity under
consideration. (Here $s,t,u$ are Mandelstam variables.)
The fact that any massless
state gives an identical contribution
up to a sign is in agreement with the supersymmetry identities [\use\Susy].

This type of calculation would be exceedingly difficult with
conventional techniques, given the complexity of the gravity three-
and four-point field theory vertices.  This may be compared to the
string-based technique where the calculation of the above helicity
amplitude is reduced to an elementary exercise.  It is amusing that
the string-based gravity calculation is only slightly more difficult
than the gluon calculation.  It is intriguing that in terms of
conventional field theory the required reorganization to obtain this
simplicity is fairly difficult to guess without some input from string
theory.

\vskip .2 cm
\noindent {\bf 8. A Conjecture for an Arbitrary Number of External Legs.}
\vskip .1 cm

It is possible to construct a conjecture for the $n$-point all plus
gluon helicity amplitude [\use\AllN], using properties of the amplitude as two
legs become collinear, and using the explicitly calculated five-point
amplitude $A_{5;1}(1^+,2^+,3^+,4^+,5^+)$ to jump-start the
conjecture.  This conjecture is analogous to the
one for maximal helicity violating tree-amplitudes
formulated by Parke and Taylor [\use\ParkeTaylor] and
proven by Berends and Giele [\use\Recursive].

The conjecture for the gluon contribution to the loop is given by
$$
 A_{n;1}(1^+,2^+,\ldots,n^+)\ =\ {i\over 96\pi^2}\,
   { E_n + O_n \over \spa1.2 \spa2.3 \cdots \spa{n}.1 }\ ,
\eqn\allnplus
$$
where the parity-odd terms are given by
$$
  O_n\ =\ 4 i \sum_{1\leq i_1 < i_2 < i_3 < i_4 \leq n-1}
            \pol_{\mu\nu\rho\sigma} k_{i_1}^\mu k_{i_2}^\nu
                                    k_{i_3}^\rho k_{i_4}^\sigma .
\eqn\oddansatz
$$
To describe the even terms define
$$
\eqalign{
  t^{(p)}_i
    \ &=\ (k_i+k_{i+1}+\cdots+k_{i+p-1})^2\ ,\cr
  t^{(2)}_i\ &=\ (k_i + k_{i+1})^2\ =\ s_{i,i+1}\ ,\cr
  t^{(1)}_{i}\ &=\ 0\ ,\cr }
	\eqn\tdef
	$$
and with all indices mod $n$:
$$
\eqalign{
  X_{p,p}^{(n)}\ &=\ \sum_{i=1}^n  t^{(p)}_i
    t^{(p)}_{i+1} \ , \cr
  X_{p-1,p+1}^{(n)}\ &=\ \sum_{i=1}^n  t^{(p-1)}_{i+1}
    t^{(p+1)}_{i} \ . \cr}
\eqn\xppdef
$$
Then the ans\"atze for odd and even $n$ are
$$
\eqalign{
  E_{2m+1}\ &=\ \sum_{p=2}^m
   \left( X_{p,p}^{(2m+1)} - X_{p-1,p+1}^{(2m+1)} \right)\ , \cr
  E_{2m}\ &=\ \sum_{p=2}^{m-1}
   \left( X_{p,p}^{(2m)} - X_{p-1,p+1}^{(2m)} \right)
  \ +\ {1\over2} \left( X_{m,m}^{(2m)} - X_{m-1,m+1}^{(2m)} \right)
  \ . \cr }
\eqn\evenansatz
$$
Note that $X_{1,3}^{(n)} = 0$, since $t^{(1)}_i = 0$.  The expression
(\use\allnplus) is cyclicly symmetric, and in the limit that two legs
become collinear, is proportional to the corresponding $(n-1)$-point
amplitude.  This conjecture will be discussed more fully elsewhere
[\use\AllN].  (A very recent paper by Mahlon gives the result for
the corresponding all-$n$ expression in QED [\use\Mahlon].)

\vskip .1 cm
\noindent{\bf 9. Summary and Conclusions}
\vskip .2 cm

In this talk, we reviewed the string-based method for evaluating
one-loop $n$-gluon amplitudes [\use\Long].  The method was originally
derived by taking the field theory limit of an appropriately
constructed [\use\Beta] four-dimensional string theory [\use\KLT].
Using the string-based method, together with a simple integration
method [\use\Integrals] and improvements in the spinor helicity
method, a first calculation of all five-gluon helicity amplitudes has
been performed [\use\AmplLet].  The five-point amplitude can be used
to jump-start a conjecture for the all plus helicity amplitude with an
arbitrary number of external legs [\use\AllN].  Using a direct string
approach we have also calculated four-point helicity amplitudes with
two external quarks and gluons [\use\Unpublished] which agree with
ref.~[\use\Kunszt].  The first calculation of a one-loop four-point
graviton amplitude has also been performed in a direct string theory
approach [\use\Gravity].

The string based rules have been reinterpreted in terms of field
theory in refs.~[\use\Mapping,\use\Matt].  The field theory ideas
obtained in this way can then be applied more widely to a variety of
problems to aid in computations. This was used as an aid in the
calculation of five-point one-loop amplitudes with external gluons and
quarks [\use\Fermion].  Another example is the application of these
ideas to certain weak interaction processes to significantly improve
their calculational efficiency [\use\Weak].  Some progress has also
been made on the extension of string-based methods to multiloops
[\use\Roland].

In summary, string theory has been useful for calculations of
gauge theory amplitudes required by experiments.

\vskip .2 cm

The work of ZB was supported in part by the US Department of Energy
Grant DE-FG03-91ER40662 and in part by the Alfred P. Sloan Foundation
Grant BR-3222.  The work of LD was supported by the Department of
Energy under contract DE-AC03-76SF00515.  The work of DAK was
supported by the {\it Direction des Sciences de la Mati\`ere\/} of the
{\it Commissariat \`a l'Energie Atomique\/} of France.

\nobreak
\immediate\closeout\rfile%\parindent=20pt
   \vskip .2 cm \noindent{{\bf References}}\vskip .1 cm \frenchspacing%
    \input refs.tmp\vfill\eject\nonfrenchspacing

\bye